# Synthesis and Characterization of Excess Magnesium $MgB_2$ Superconductor under Inert Carbon Environment

**B. B. Sinha[a], M. B. Kadam[a], M. Mudgel[b,c], V. P. S. Awana[b,*], Hari Kishan[b] and S. H. Pawar[a,d]**

[a]Superconductivity Lab, Department of Physics, Shivaji University, Kolhapur 416 004 Maharashtra, India
[b]National Physical Laboratory, Dr. K. S. Krishnan Marg, New Delhi 110012, India
[c]Deptartment of Physics and Astrophysics, Delhi University, New Delhi-110007, India
[d]Department of Technology, D. Y. Patil University, Kolhapur 416006

**Abstract:**

The structural, transport and magnetic properties of $MgB_2$ superconductor heavily blended with Mg is studied. The samples are synthesized with a new approach in both, pressed carbon environment and in flowing argon. The excess magnesium used is observed to play dual role: one being the prevention of Mg losses during the synthesis process and hence maintaining the stoichiometry of $MgB_2$ phase, and second being the formation of Mg milieu probably all around the $MgB_2$ grains to give a ductile and dense structure. Excess Mg also improves the grain connectivity by going in to the pores and there by minimizing the insulating junctions. The residual resistivity of the sample is observed to decrease from 57.02 μΩ cm to 10.042 μΩ cm as it is progressively filled with superconductor-normal-superconductor (SNS) type junctions amongst the grains by the virtue of increased magnesium content. The synthesized samples devoid of porosity show the superconducting transition, $T_c$ in the range of 39K – 34K as of clean $MgB_2$ samples, though overloaded with Mg. The excess Mg resulted in enhanced critical current density, $J_c$ from $6.8x10^3$ A $cm^{-2}$ to $5.9x10^4$ A $cm^{-2}$ at 20K and 10KOe, with reasonable decrease in the superconducting transition. Thus our samples being overloaded with Mg impart mechanical strength and competitive superconducting properties, which forms a part of interest.

*Corresponding Author:

Dr. V.P.S. Awana

Fax No. 0091-11-45609310: Phone no. 0091-11-45609210

e-mail-awana@mail.nplindia.ernet.in: www.freewebs.com/vpsawana/

## Introduction:

The $MgB_2$ superconductor with a highest $T_c$ of 39 K [1] and multiple gap [2, 3] superconductivity is one of the interesting superconductors. Besides, it is a BCS superconductor [4] having $T_c$ with in strong coupling BCS limit [5]. Any attempts to increase the superconducting transition temperature [6-9] have not achieved success yet [10-12], due to its typical phonon coupling mechanism [13-18]. But the other critical parameters like critical current density, and upper critical field can be enhanced remarkably by nano- particle doping [19-26]. Moreover, the mechanical properties of $MgB_2$ superconductor also vary depending upon the dopant as well as on the synthesis route. The synthesis route partially dictates the density of the sample and the impurity level influencing the properties of $MgB_2$ phase. The synthesis technique for $MgB_2$ superconductor so far needed a pressurized environment to achieve better superconducting properties. The samples synthesized under pressure are comparatively dense [27] than those synthesized in ambient environment [28] and hence have high superconducting homogeneity with reduced weak links and subsequently possess high critical current density [29].

In present work, instead of expensive nano particles we introduce a cheaper way to enhance critical current density of $MgB_2$ through heavy blending of Mg in $MgB_2$ as against mere doping. Also the blending of Mg may result in better ductile strength [30,31], which is good for fabrication of wires/tapes. Secondly, the samples are synthesized by a new versatile easy technique in a pressed carbon environment under flowing argon avoiding expensive HPHT (high pressure high temperature) method. The effect of excess Mg on the transport as well as magnetic properties of $MgB_2$ is studied systematically. The enhancement of critical current density is observed with increasing content of Mg.

## Experimental

The solid-state reaction route was followed for the synthesis of superconducting $MgB_2$ samples. The primary constituents used were 99% pure Mg powder and 99% pure amorphous boron powder. The components were blended using agate mortar with excess of Mg for about 3 hours. Both stoichiometric and off-stoichiometric samples were prepared by varying Mg and B content in different Mg:B ratio of 1:2 (stoichiometric) , 1:1 (double the amount of Mg as compared to stoichiometry) and 1.5:1 (triple the amount of Mg as compared to stoichiometry) to give samples coded as M1, M2, and M3 respectively. The blend with these different ratios of Mg:B was transformed into pellet form under the pressure of 10 tons/cm$^2$. The pellet was then placed in a stainless steel tube with length of about 1 foot, near the closed end and embedded in carbon powder which was filled along the whole length of the tube and pressed in a tightly packed manner to form a column of carbon powder. This column of carbon would act as a simple barricade to prevent the loss of Mg and sustain Mg pressure in assembly. The pellet was covered with Mg powder, being used as a spacer in between carbon powder and the pellet. The whole tube was placed into tube furnace in a position such that the sample reaches at the core of the furnace so as to have a sufficient temperature gradient along the length of carbon column. The tube of the furnace was evacuated and argon was flushed during the heat treatment to remove the remnant air or oxygen from the environment. Once the argon was flushed, the sample was heated at 900$^o$C for about 3 hours. The highly dense samples obtained after heat treatment were tested for their structure by XRD measurement studies using Cu Kα radiations in the 2θ range of 20-80°. The growth morphology of the sample was studied by scanning electron microscopy (SEM) JEOL JSM 6360. The samples were further tested for its superconductivity by resistivity meausurements through standard four-probe technique, using atomized APD closed cycle refrigerating system. The superconductivity in the sample was also confirmed through field cooled and zero field cooled measurement. The critical current density of the samples was extracted from the plot of magnetization with respect to field using Beans critical state model [32] considering the sample as a whole responsible for critical current density as against individual grains. The magnetization measurements were done with a quantum Design SQUID magnetometer (MPMS-XL). The sample dimensions used for the measurements were more or less rod type. The flux pinning force was calculated and was studied for the effect of excess magnesium used in the synthesis of the $MgB_2$ samples.

## Result and Discussion:

*(i) X-ray diffraction studies:*

The samples prepared by the proposed synthesis technique were subjected to X-ray diffraction analysis to confirm its structure. The powder XRD patterns for all three samples M1, M2 and M3 with the Mg:B ratio of 1:2, 1:1 and 1.5:1 are as shown in the figure 1 (a-c) respectively. As seen in the figures the main phase for all the samples is $MgB_2$ crystallized in hexagonal structure. None of the phase consisting of carbon was detected which indicates that the samples were well insulated from carbon powder used. The Mg metal forms the next major phase present in the sample. As the amount of Mg in the sample increases, the intensity for the Mg phase peak increases. The calculated lattice parameters for all the samples are as given in Table 1. It is observed that with increase in Mg content the inplane *a-axis* lattice parameter decreases while the inter plane *c-axis* lattice parameter increases. This leads to an ultimate increase in the aspect ratio ($c/a$) of the $MgB_2$ structure in the sample. Also its remarkable to note that with increase in Mg content, the samples become increasingly textured along the (100) direction in comparison with any other planes. This may be due to infiltration of Mg in to the $MgB_2$ lattice, which resulted in merely modified crystal structure. No other intermediate phase between Mg and $MgB_2$ is observed in the sample. There is neither $BO_x$ insulating layer nor the formation of $MgB_4$ insulating phases as mentioned in Ref. [33], occurs at the grain boundaries. Only an extra peak of MgO is noticed which may have formed due to traces of oxygen present in the ambiance. The intensity of MgO peak increases with the increasing Mg content.

*(ii)Morphological Studies:*

Figure 2 (a, c & e) shows the morphology of M1, M2 and M3 samples respectively at the magnification of 5000x. The surface morphology of the samples appears to become rough with the increase in Mg content. Moreover it is remarkable to note that even though the amount of Mg in the sample is increased, there is hardly any porosity observed in present samples. $MgB_2$ samples are usually porous due to high vapor pressure of Mg [34]. To have a more clear understanding the high-resolution morphology of the samples is given [Figure 2(b, d & f)], which also reveals pores free morphology of samples even in the case of excess Mg samples. This indicates that the excess Mg used have not ventured in to $MgB_2$ lattice beyond the solubility limit but have separated out itself from the $MgB_2$ grain to form $MgB_2$/Mg matrix. Moreover, MgO formed during synthesis also may enhance the grain connectivity and core density of $MgB_2$ matrix. It could also cause the grain boundary pinning and contribute towards the enhancement of critical current density [35].

*(iii)Electrical transport studies:*

The electrical transport measurements with respect to decrease in temperature for the samples M1, M2 and M3 are shown in the figure 3. It is seen that the superconducting transition for sample M1 was observed at 38.67 K while the samples M2 and M3 showed the transitions at about 35.25K and 34.38K respectively. Sample M1 with stoichiometric proportion of Mg and B showed an exponential temperature dependence of resistivity while other samples M2 and M3 with excess Mg showed linear temperature dependence of resistivity up to the superconducting transition. Superconducting transition in the sample M1 appears to be quite sharp while as the amount of Mg increases the transition width increases for M2 and M3. It is interesting to note that the residual resistivity ρ(40K) for the samples decreases with the increase in the Mg content but ρ(300K) does not change in a regular fashion.

It is known that the inclusion of impurity in $MgB_2$ lattice reduces the $T_c$ of the sample [36-38] due to scattering effects [39-41]. Here, though sample M1 consists of Mg as depicted in XRD, the transition temperature observed is not reduced and is quite sharp. This implies that whatever Mg present, is at grain boundaries as pores eliminators and has not infiltrated in to $MgB_2$ lattice as an impurity. This is in agreement with the fact that the crystal structure of sample M1 remained structurally undistorted. As the amount of Mg is increased for sample M2 and M3, transition temperature gradually gets decreased to 35.25K and 34.38K respectively. This indicates that though the Mg remains at the grain boundaries, if excess, it infiltrates in to the $MgB_2$ grains to a certain solubility limit to decrease $T_c$ by making the $MgB_2$ grains less clean though not dirty [2]. Though the amount of Mg is considerably increased (doubled for M2 and tripled for M3) in present case as compared to reports earlier [31, 42], the transition temperature still remains more or less competitive if compared. This strongly supports the fact that Mg goes into the $MgB_2$ lattice only up to certain solubility limit and beyond that accumulates around the $MgB_2$ grain boundaries to make SNS type junctions. This can be further supported by the observed temperature dependence of resistivity for samples M2 and M3 which becomes linear as compared to exponential behaviour of sample M1, with increase in Mg content. The residual resistivity ρ(40K) values are observed to decrease with increased amount of Mg, which may be due to the formation of low resistive phases of Mg (1.6 μΩ cm at 300K), which provides current percolation, along with the increase in the density of the sample due to Mg accumulation. It also appears that though the transition temperature for samples M2 and M3 remains more or less same, the transition width for the sample M3 is increased remarkably as compared to that of sample M2. The increased broadening in the superconducting transition for sample M3 is accompanied with the drastic decrease in the overall resistivity. This is due to the enhanced formation of nothing but dense Mg metal matrix in the sample, which surrounds the superconducting $MgB_2$ grains all over the sample. The increased RRR($\rho_{300}/\rho_{40}$) value in case of M2 and M3 samples is also due to the Excess Mg present at the grain boundaries so that current find a less resistive path and relatively more slope is observed in the R-T curve.

*(iv)DC susceptibility measurements:*

Figure 4 shows the zero field cooled (ZFC) and field cooled (FC) plots at 10 Oe for the samples M1 and M3 which forms the two extreme samples with the ratios 1:2 and 1.5:1 respectively. It is observed that for both the samples ZFC branch shows a sharp one-step diamagnetic transition at 37.5K and 35.5K respectively for M1 and M3 samples. Thus the transition temperature decreases with the increase in the Mg content, which is in agreement with the transitions observed in resistivity measurements. The sample M3 with more Mg content showed somewhat rounded nature after superconducting transition before it attains the saturation down to 5K; the same is not as prominent as in sample M1. Moreover the extent of diamagnetic signal decreases with increased Mg content. It is interesting to notice FC plots where both the samples show paramagnetic Meissner effect (PME) [in confirmation with Ref. 43, 44] with the transition at same temperature as that of ZFC plot. This confirms the formation of SNS type of junctions [43,45], which leads to trapped flux and weak FC signal [46]. This is in agreement with the observed XRD in which extra phase of Mg was observed which takes part in the formation of SNS junctions to give rise to PME. The presence of SNS junctions in the samples prove the formation of Mg metal matrix around the superconducting $MgB_2$ grains.

*(v)Magnetization measurements:*

The magnetic hysteresis behavior (M-H) for all the samples M1, M2 and M3 at 5K, 10K and 20K is shown in Figure 5 under the applied field up to 70 kOe. For stoichiometric sample M1, the M-H loop closes much before than the excess Mg samples M2 and M3 at all measured temperature of 5, 10 and 20K. Quantitatively at 20K, the M-H loop closes at about 30 kOe for sample M1 while it remains open above 40 kOe for M3. Thus, additions of Mg apparently increase the irreversibility field, $H_{irr}$ values of the samples.

The $J_c$ values were calculated from these M(H) response by revisiting the beans model [32] with the sample dimensions . The $J_c$ plots for M1, M2 and M3 samples at 20K are shown in the Figure 6. Inset shows the same at 20 K. It reveals that the critical current density for the samples with increased Mg content is improved considerably as compared to the stoichiometric M1 sample. At the temperature of 20K, the critical current density value for the sample M1 comes out to be $6.8x10^3$ A $cm^{-2}$ at 10 kOe while the same is enhanced to $3.4x10^4$ A cm-$^2$ and $5.9x10^4$ A $cm^{-2}$ for the Mg added M2 and M3 samples respectively. Thus $J_c$ increases by an order of magnitude with Mg doping in stoichiometric $MgB_2$. The similar behavior of increase in $J_c$ is also noticed at lower temperatures of 5 and 10K. Thus we come to the conclusion that although excess Mg results into decrease in transition temperature but at the same time it successfully acts as a pinning centre; introduces some other pinning centres like MgO and in turn results into significant enhancement of $H_{irr}$ and $J_c$(H) values. The observed values of critical parameters for stoichiometric sample M1 are competitive with the literature [47,48] and we have introduced a way to enhance it's superconducting performance just by varying the stoichiometry without using any foreign particle. Thus the Mg doping seems to be a cheaper and easier way in comparison to the costly nano particle doped $MgB_2$ samples.

The flux pinning behavior of the samples at T=10K is shown in Figure 7. The relationship between flux pinning force and critical current density could be described by [49-51],

$$F_p = \mu_o J_c(H) H$$

where $\mu_o$ is the magnetic permeability in vacuum. Main panel in Figure 7 show the variation of reduced flux pinning force with field while the inset plots total flux pinning force. In both, the flux pinning plots for sample M2 is shifted slightly towards higher field as compared to the sample M1 while M3 shows a profound shift towards higher field. This indicates improvement in the flux pinning forces for samples M2 and M3 as compared to that of sample M1 in confirmation with the $J_c$ (H) results. As far as the type of pinning is concerned, it is the combination of grain boundary pinning due to Mg and MgO present outside the lattice and the pinning due to point defects, which arise from Mg inclusion in $MgB_2$ matrix [50]. The fact is supported by the behavior of Flux pinning plots for samples M2 and M3 in which the peak is not only broadened but also gets shifted towards right say by about 1 Tesla in case of M3. Hence the flux pinning plots again support the fact that the Mg added into the samples acts as an effective flux pinning centers, which results in the improvement of critical current density of the sample.

**Conclusions:**

The superconducting properties of the $MgB_2$ sample heavily loaded with Mg having the Mg:B ratio or 1:2, 1:1 and 1.5:1 are studied successfully. It is remarkable to observe that though the sample is heavily loaded with Mg, the sample speculated to have maximum Mg concentration showed the superconducting transition at about 34 K with a mere decrease in transition temperature of about 5 K. The addition of Mg in the sample resulted in overall decrease in the resistivity of the sample, which is due to accumulation of Mg around the $MgB_2$ grains to form a superconductor-metal matrix and simultaneously improving the overall density of the samples. The enhancement in the irreversibility field was observed from the magnetization curves with the increase in Mg content. The overall presence of Mg metal matrix in the sample enhanced the flux pinning properties with increase in pinning centers. This in turn enhanced the critical current density of the samples from $6.8x10^3$ A $cm^{-2}$ to $5.9x10^4$ A $cm^{-2}$ at 20K and 10 kOe. Hence it is interesting to have such a sample with enhanced critical parameters just by varying the stoichiometry of the sample, which could be considered for the fabrication of wire or tapes for various applications.

**Acknowledgement**

The authors from *NPL* would like to thank Dr. Vikram Kumar (*DNPL*) for his great interest in present work. Prof. E. Takayama muromachi is acknowledged for the magnetization measurements at 7Tesla- SQUID magnetometer (MPMS-XL). M. Mudgel and B. B. Sinha would like to thank the *CSIR* for the award of Senior Research Fellowship to pursue their *Ph. D* degree.

## Figure captions

Figure 1. XRD pattern for different samples with Mg:B ratio of (a) 1:2, (b) 1:1 (c) 1.5:1

Figure 2. SEM picture with two different magnifications 5000 x and 10000 x respectively of samples M1 (a) & (b), M2 (c) & (d) and M3 (e) & (f).

Figure 3. Resistivity plots for samples M1, M2 and M3 with different Mg:B ratio of 1:2, 1:1 and 1.5:1 respectively. Inset shows the transition zone of all samples in enlarged view.

Figure 4. Field cooled and Zero field cooled plots for the samples M1 and M3.

Figure 5. Magnetization plots for the samples M1, M2 and M3 at three different temperatures of 5K, 10K and 20K.

Figure 6. Critical current density plots for the samples M1, M2 and M3 at (a) T=20K, the inset shows the same at 10K.

Figure 8. Reduced Flux pinning plots for the samples M1, M2 and M3 at T=10K, inset shows the variation of unnormalized total flux pinning force with field.

Table1. Lattice parameters and *c/a* ratio for the samples M1, M2 and M3

| Sample | a(Å) | c(Å) | *c/a* |
|---|---|---|---|
| M1 | 3.0838 | 3.539 | 1.147 |
| M2 | 3.0754 | 3.5547 | 1.155 |
| M3 | 3.0727 | 3.5534 | 1.156 |

Fig. 1

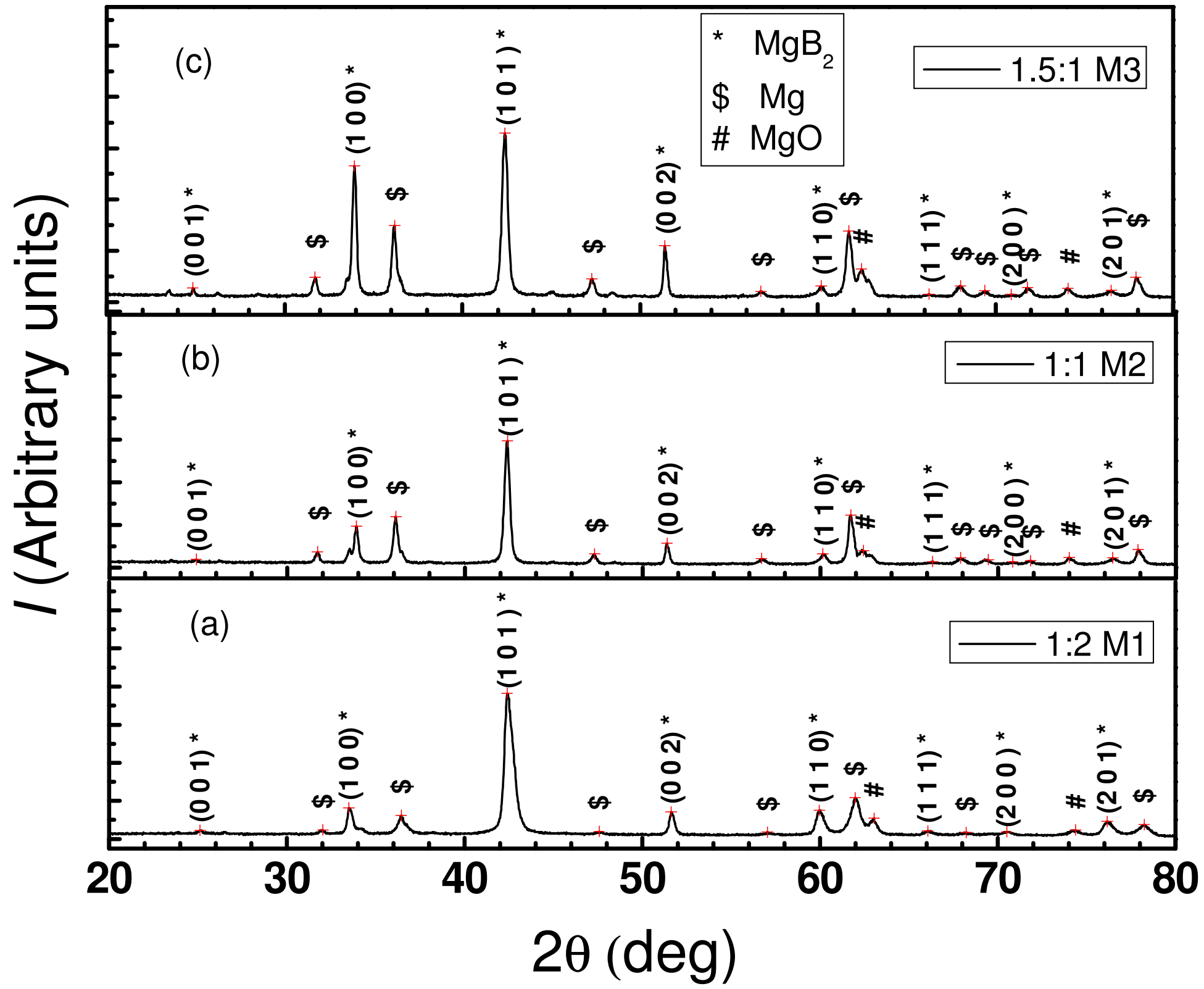

Fig. 2

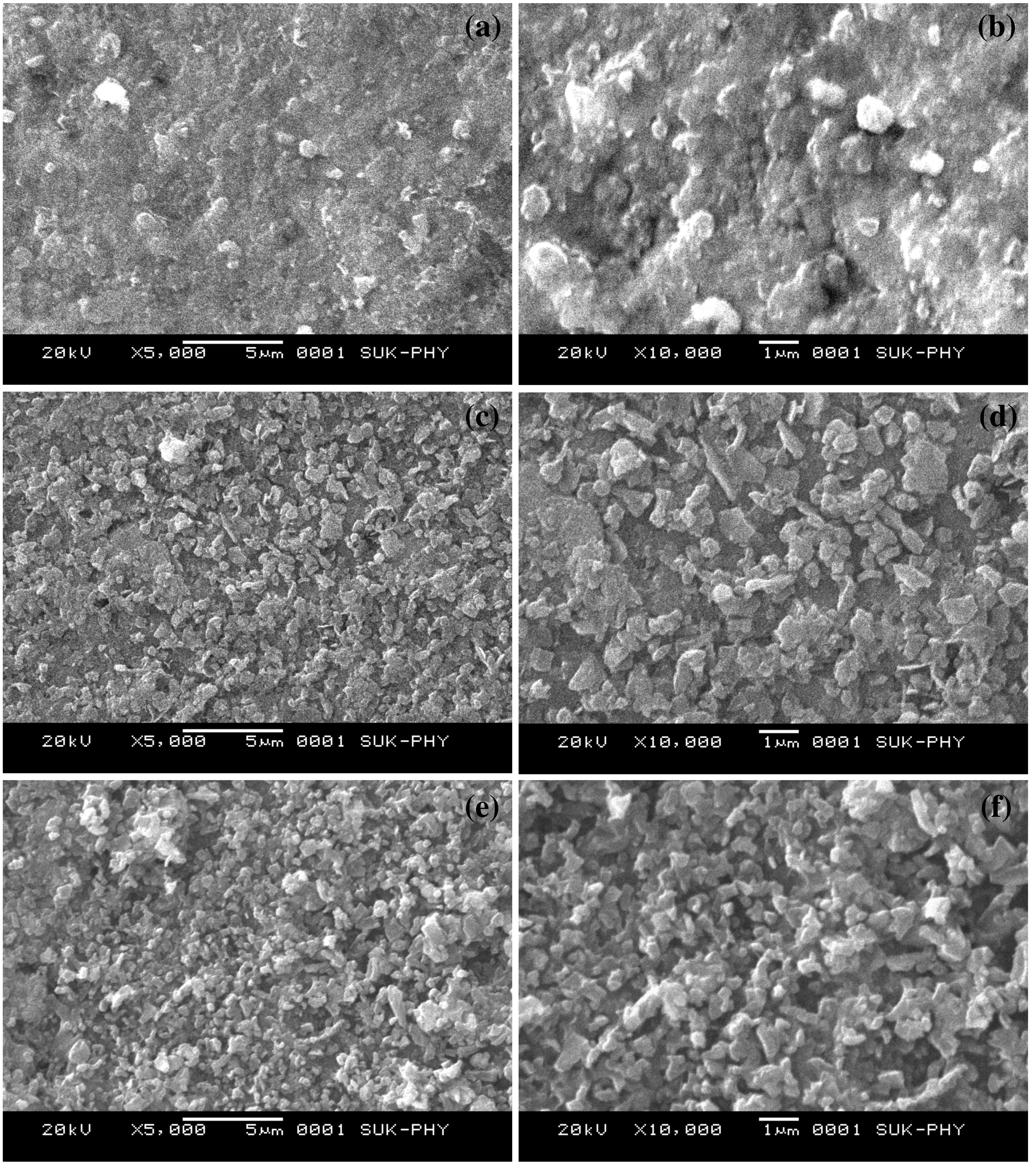

Fig. 3

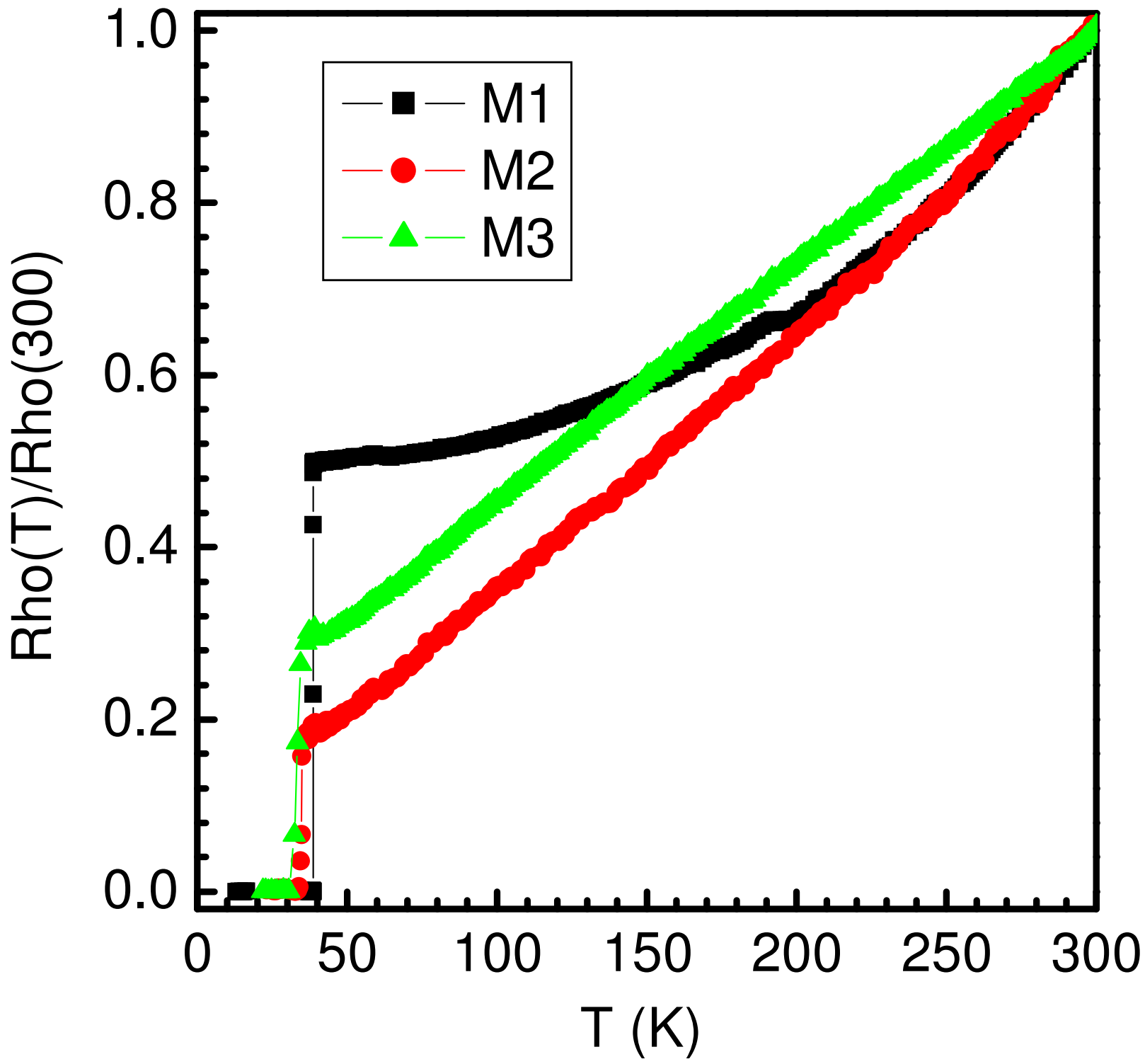

Fig. 4

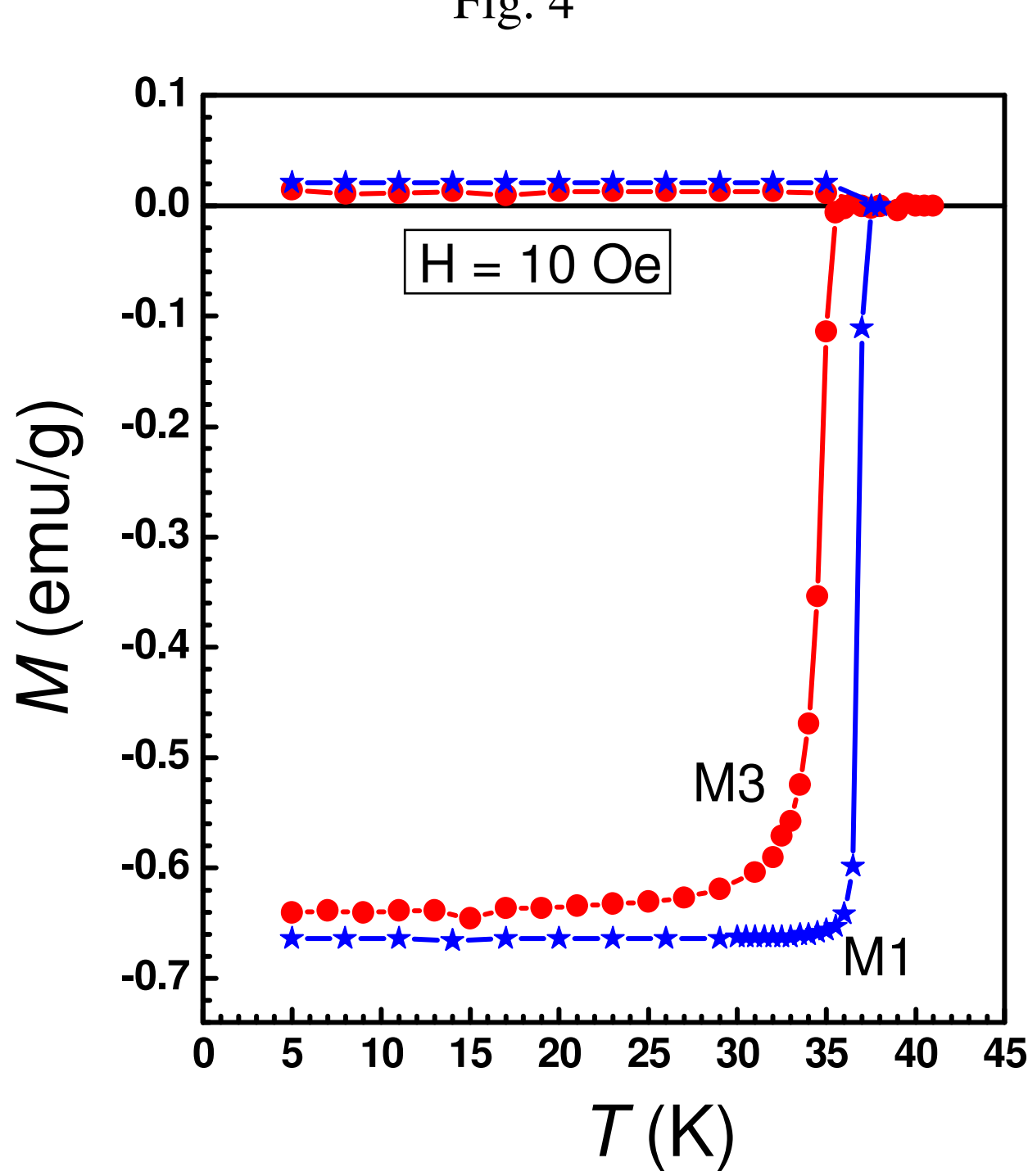

Fig. 5

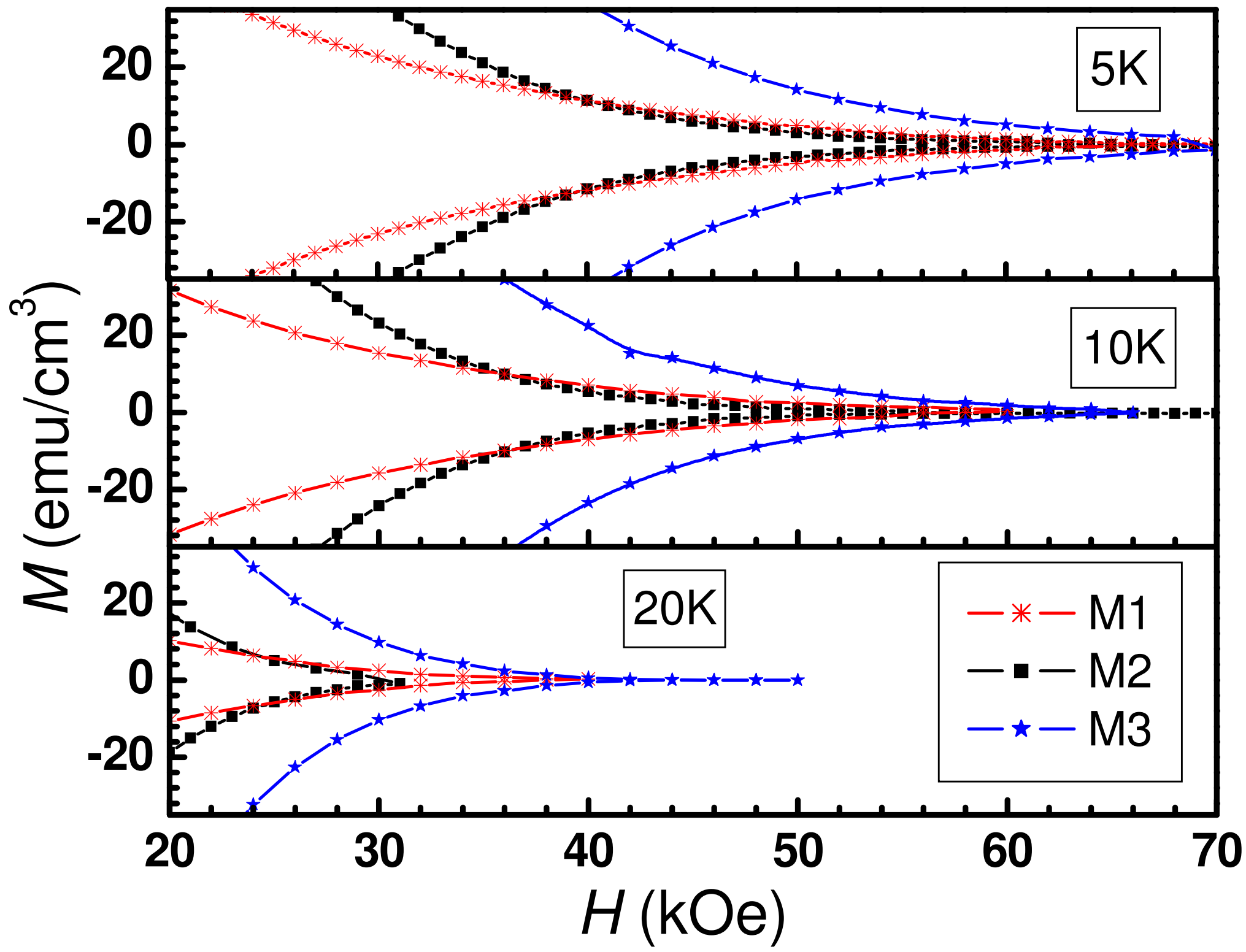

Fig. 6

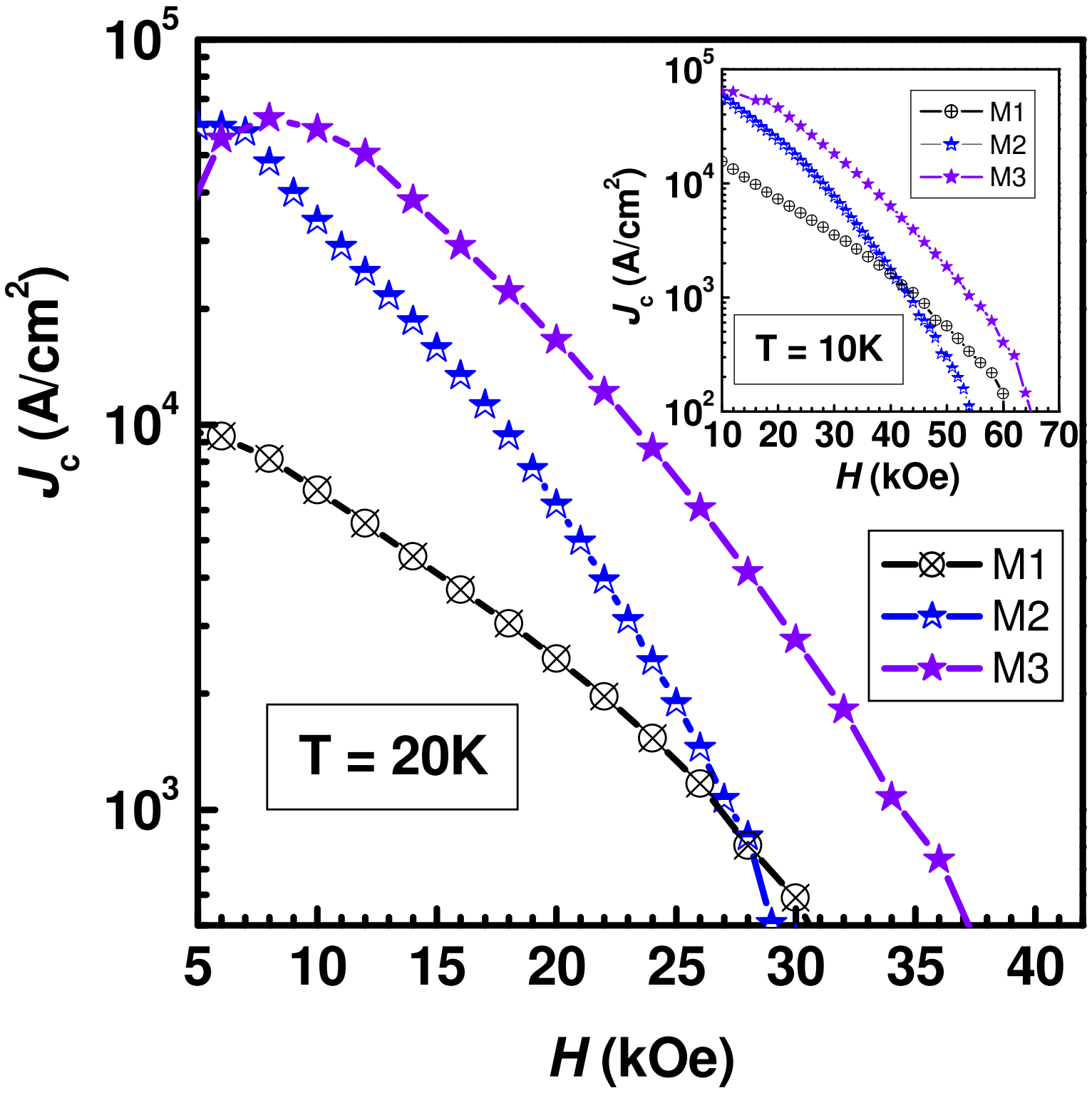

Fig. 7

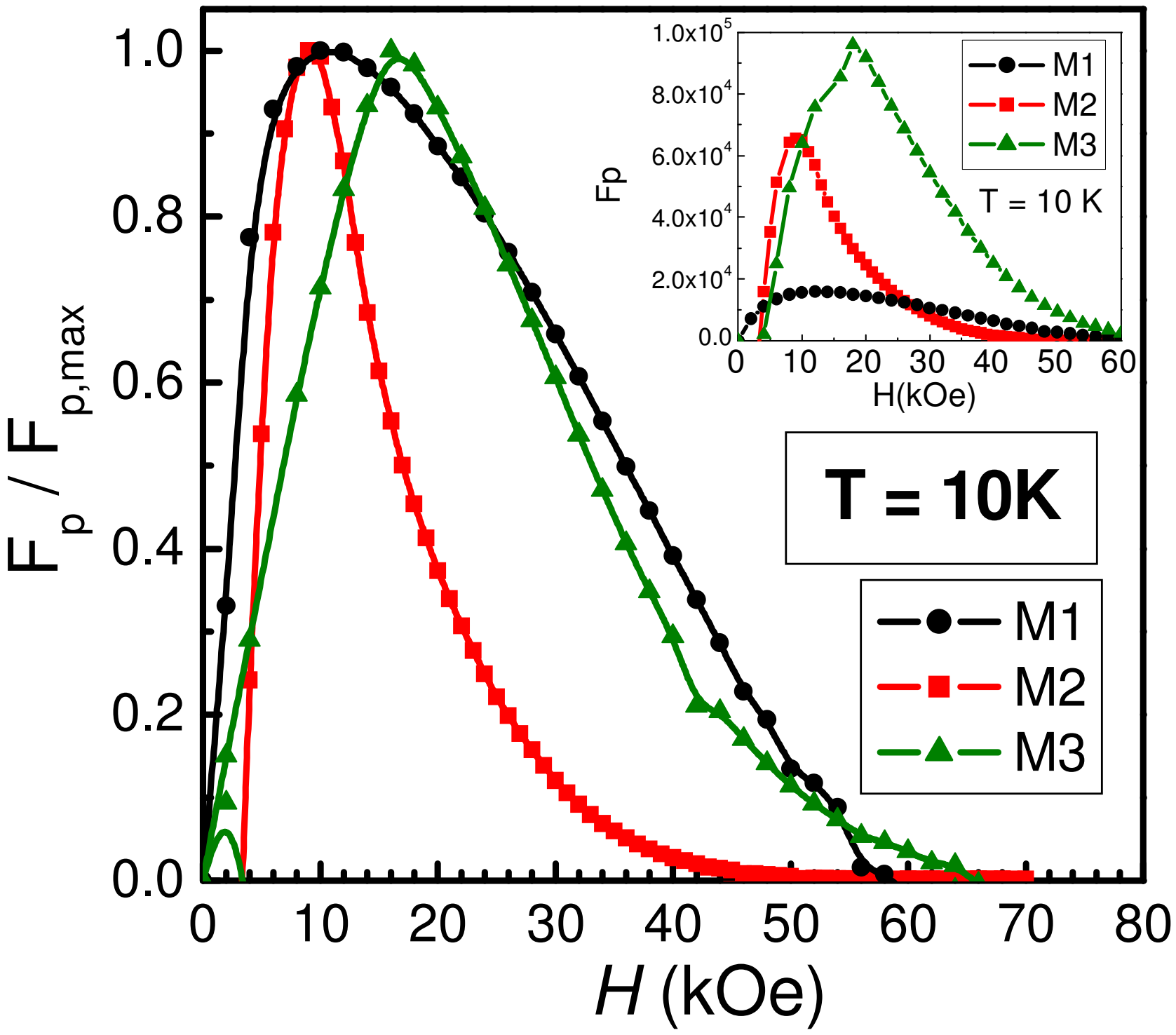